\pdfoutput=1% forces arxiv to use pdflatex to process pdf plots & figures
\documentclass[a4paper,12pt]{article}
\usepackage{ifpdf}
\ifpdf
\usepackage[pdftex]{graphicx}
\usepackage[usenames,dvipsnames]{color}
\else
\usepackage{graphicx}
\usepackage[usenames,dvips]{color}
\fi
\usepackage{amsmath}
\usepackage{amssymb}
\usepackage[colorlinks=true]{hyperref}   % for electronic versions
\usepackage{caption}
\usepackage{paper}

\numberwithin{equation}{section}

\begin{document}

\title{An Einstein-Bianchi system for\\[5pt]
       Smooth Lattice General Relativity. II.\\[5pt]
       3+1 vacuum spacetimes.}
\author{%
Leo Brewin\\[10pt]%
School of Mathematical Sciences\\%
Monash University, 3800\\%
Australia}
\date{29-Jan-2011}% 1st version started
\date{22-Mar-2011}% 1st version finished
% \date{08-Jan-2010}% finished
% \date{02-Nov-2009}% in print
\reference{Preprint}
% \reference{Preprint: arXiv:0903.5365\\[5pt]
%            Journal: {\it Phys.Rev.D.} {\bf 80} 084030 (2009)}

\maketitle

\begin{abstract}
\noindent
We will present a complete set of equations, in the form of an Einstein-Bianchi system, that
describe the evolution of generic smooth lattices in spacetime. All 20 independent Riemann
curvatures will be evolved in parallel with the leg-lengths of the lattice. We will show that
the evolution equations for the curvatures forms a hyperbolic system and that the associated
constraints are preserved. This work is a generalisation of our previous paper
\cite{brewin:2010-02} on the Einstein-Bianchi system for the Schwarzschild spacetime to
general 3+1 vacuum spacetimes.
\end{abstract}

% ============================================================================================
\section{Introduction}
\label{sec:intro}

In a series of papers we have shown that the smooth lattice method works remarkably well
for simple spacetimes such as %
the Schwarzschild spacetime in various slicings \cite{brewin:2010-02,brewin:2002-01}, %
the maximally sliced Oppenheimer-Snyder spacetime \cite{brewin:2009-05}, %
the vacuum Kasner cosmologies \cite{brewin:1998-02} and for constructing %
Schwarzschild initial data \cite{brewin:1998-01}. %
The equations are simple and require little computational sophistication to achieve stable and
accurate results. The real test of the method however must be in the context of generic
spacetimes. This paper is a first step in that direction.

The logic behind the smooth lattice approach is quite simple. We assume that we are given a
lattice, built from a large collection of interconnected vertices, and where each path that
connects a pair of vertices is taken to be a geodesic segment of the spacetime. The only data
that we are given for the lattice is the connection matrix (which describes the topology as a
list of pairs of connected vertices) and the lengths of each geodesic segment (which describes
the metric properties). In this picture we are assuming that the lattice geometry is a close
approximation to some underlying smooth geometry. The question that (should) spring to mind is
-- Given the leg lengths on a lattice, how do we compute the Riemann curvatures? We will
return to this important question in just a moment, but for now let us suppose we have a
suitable algorithm by which we can accurately compute the Riemann curvatures. It is then a
simple matter to impose the vacuum Einstein equations\footnote{For pure pedagogy we will
restrict the discussion to vacuum spacetimes.} which in turn will impose
constraints\footnote{Not to be confused with any constraints that may exist at the continuum
level, for example the ADM constraints.} on the leg-lengths. This furnishes us with a discrete
set of equations for the leg-lengths. Solving these equations will yield a discrete solution
of the vacuum Einstein equations.

We now return to the question of how to recover the Riemann curvatures given the set of
leg-lengths. In one of our earlier papers \cite{brewin:1998-01} we argued that if the lattice
was sufficiently well refined then a local Riemann normal coordinate frame could be
constructed in the neighbourhood of any vertex extending to include, at least, the immediate
neighbouring vertices. We called this neighbourhood the computational cell for the vertex (for
lattices built from tetrahedra this would consist of the tetrahedra attached to the vertex).
In this computational cell we can expand the metric as a power series \cite{brewin:2009-03}
around the central vertex
\begin{equation}
g_{\mu\nu}(x) = g_{\mu\nu} 
               - \frac{1}{3} \Rmanb x^\alpha x^\beta 
               + \BigO{L^3}
\label{eqn:RNCMetric}
\end{equation}
where $L$ is a typical length scale for the computational cell. The requirement that the legs
are geodesic segments leads, after some detailed calculations \cite{brewin:2009-03}, to the
following equation
\begin{equation}
\Lsqij = g_{\mu\nu} \Dxij^\mu \Dxij^\nu
            - \frac{1}{3} \Rmanb x^\mu_i x^\nu_i x^\alpha_j x^\beta_j
         + \BigO{L^5}
\label{eqn:RNCLsq}
\end{equation}
where $\Dxij^\mu:=x^\mu_j-x^\mu_i$. The approach advocated in \cite{brewin:1998-01} was to use
this equation to extract the Riemann curvatures from the lattice. This may sound simple but
there are a number of troubling issues.

The first issue concerns the coordinates. How do we compute coordinates for each vertex? Some
can be set by simple gauge transformations (\eg the origin can be tied to the central vertex)
while the remainder must be computed from the lattice data (\ie the leg-lengths). This forces
us to view the above equations (\ref{eqn:RNCLsq}) as a coupled system for the curvatures and
the coordinates.

The second issue is one of accountancy -- do we have enough equations to compute the curvatures
and the coordinates? For most lattices (in 3 and higher dimensions) the legs out number the
coordinates $x^\mu_i$ and curvatures $R_{\mu\alpha\nu\beta}$. As an example, the computational
cell used in our earlier paper \cite{brewin:1998-01} contained 78 legs and 19 vertices. Thus we
had 78 equations for 6 curvatures and 57 coordinates (of which 6 can be freely chosen). There
are at least two ways to handle this over supply of information. We can either form linear
combinations of the above equations (\ref{eqn:RNCLsq}) to produce a reduced system in which the
number of equations matches the number of unknowns. Or we can include a sufficient number of
higher order terms in the Taylor series so as to produce a consistent set of equations. This
later approach has the possible benefit of producing higher order approximations for the
$R_{\mu\alpha\nu\beta}$ but at considerable extra expense. In both instances we still have a
large coupled non-linear system of equations to solve at each vertex and at each time step.
This is a considerable computational challenge.

Another important issue is one of uniqueness -- how many distinct solutions can we find for the
$x^\mu_i$ and $\Rmanb$? The equations are non-linear and thus it is conceivable that more than
one solution could be found. Do the solutions form a continuous family or are there only a
finite set of solutions? How would we choose between these solutions? In our earlier paper
\cite{brewin:1998-01} we resolved these problems by extending the lattice data to include the
angles between each pair of legs attached to the central vertex. This allowed us to obtain an
explicit and unique solution for all of the coordinates in a computational cell. It also had
the added bonus of decoupling the coordinates from the curvatures -- we could calculate all of
the coordinates before computing the curvatures. The price we paid for this improvement was a
significant increase in the number of data to be evolved. Where previously we had 78 legs per
computational cell, now we had a further 33 angles.

However, there is a final issue which is much more serious than those just mentioned. To
obtain $\BigO{L}$ accurate estimates for the curvatures, the coordinates must be computed to
at least $\BigO{L^4}$ accuracy (\ie the errors must be no worse than $\BigO{L^4}$). This
follows by inspection of equation (\ref{eqn:RNCLsq}). Suppose the error in $x^\mu_i$ is
$\BigO{L^a}$ for some $a>0$. This error will couple with the first term on the right hand side
of (\ref{eqn:RNCLsq}) to introduce an error of $\BigO{L^{a+1}}$. But the curvature terms are
$\BigO{L^4}$ and will dominate the error term only when $a\ge4$. Admittedly this is a somewhat
naive analysis as it takes no account of the smoothness of the underlying geometry which might
ensure that various lower order terms cancel (see for example the role smoothness plays in
establishing the truncation errors in centred finite-difference approximations). But in the
absence of an explicit algorithm we are unable to demonstrate that such cancellations do
occur\footnote{Though the introduction of angles does produce an explicit algorithm its
analysis is too unwieldily to be of any use.}. The upshot is that if we persist with any of
the variations suggested above we must design a solution strategy that guarantees, without
invoking smoothness, that the errors in the coordinates are no worse than $\BigO{L^4}$.
Despite our best efforts, we have not found a reliable solution to this problem.

These issues are not altogether new nor surprising and have proved to be a niggling concern
throughout the development of the smooth lattice method. The only working solution that we
have found (there may be others) is to surrender some (or all) of the main equations
(\ref{eqn:RNCLsq}) in favour of the Bianchi identities. In all of our papers
\cite{brewin:2010-02,brewin:2002-01,brewin:2009-05,brewin:1998-02,brewin:1998-01} we used a
combination of the Bianchi identities and the geodesic deviation equation in $1+1$ spacetimes.
The results were very encouraging. This was a hybrid
scheme\footnote{The geodesic deviation equation arises as a continuum limit of the smooth
lattice equations \cite{brewin:1998-01}.} and we attributed its success to the introduction
of the Bianchi identities. This is the motivation for the present paper -- Can we use the
Bianchi identities to compute all of the Riemann curvatures in a $3+1$ spacetime? We should
emphasise that there is one important difference between what we propose here and our previous
work. In this paper we will use the full set of Bianchi identities to \emph{evolve} all 20
independent Riemann curvatures. In contrast, in our $1+1$ experiments we used one Bianchi
identity to compute one spatial curvature (\ie a purely 3-dimensional computation within one
Cauchy surface).

Why should we believe that this use of the Bianchi identities will overcome the issues
described above? Simply, it allows us to use lower order approximations for the vertex
coordinates (even flat space approximations) without compromising the quality of the estimates
of the curvatures. We will return to this point after we have presented the full set of
evolution equations.

% ============================================================================================
\section{Notation}

A typical computational cell will be denoted by $\Omega$. This will be a compact subset of the
spacetime manifold. The central vertex of the cell will be denoted by $O$ and the subset of
$\Omega$ obtained by the intersection of $\Omega$ with the particular Cauchy surface that
contains $O$ will be denoted by $\omega$. We will describe $\omega$ as the floor of
$\Omega$. As $\Omega$ has a finite extent there will be an image of $\omega$ that defines the
future end of $\Omega$. We will refer to this as the roof of $\Omega$. We will have little to
reason to refer to the past end of $\Omega$ but calling it the basement seems consistent.

We will assume throughout this paper that the vertex world lines are normal
to the Cauchy surfaces (\ie zero drift, in the language of \cite{brewin:1998-02}). This may
seem restrictive but in our experiments to date it has worked very well.

Within $\Omega$ we will employ two sets of vectors essential to the evolution of the lattice.
The first set will be an orthonormal tetrad, denoted by $e_a$, $a=1,2,3,4$, tied to the world
line of $O$ and aligned so that $e_1$ is the tangent vector to the world line of $O$. As we
have assumed that the drift vector is everywhere zero this also ensures that $e_1$ is the
future pointing unit normal to $\omega$ at $O$. Following convention, we will write $n^\mu$ as
the unit normal to $\omega$ though as just noted, this is identical to $e_1$. The second set
of vectors will be based on the set of radial legs attached to $O$. Each leg will be of the
form $(oi)$ and we will use $v_i$ to denote the vector that joins $(o)$ to $(i)$. Note that
the $v_i$ are neither unit nor orthogonal. Latin characters will always be used to denote
tetrad indices while the spacetime indices will be denoted by Greek letters. Latin characters
will also be used as vertex labels and where confusion might arise we will use subsets of the
Latin alphabet with $a,b,c,\cdots h$ reserved for frame components while $i,j,k,l,m$ will be
reserved for vertex labels. Obviously this distinction will only be imposed for equations that
contain both types of index.

Each cell will carry a Riemann normal coordinate frame (an RNC frame), with coordinates
$x^\mu=(t,x,y,z)$, tied to the central vertex and aligned with the tetrad. Note that this
gives precedence to the tetrad over the coordinates. Coordinate components will be written as
$R_{\mu\nu}$ or for specific components as, for example, $R_{tx}$ while for frame components
we will use scripts characters ${\cal R}_{ab}$. The coordinates for a typical vertex $(i)$
will often be written as $x^\mu_i$ but on occasion we will have need to talk about the
particular values for the $x^\mu_i$ in which case we will write $(t,x,y,z)_i$ or even $x^t_i$,
$x^z_i$ etc.

Each RNC frame will be chosen so that at $O$ the metric is diagonal, $(\gmn)_o =
\diag(-1,1,1,1)$. Both spacetime and tetrad indices will be raised and
lowered, at $O$, using the metric $\diag(-1,1,1,1)$. With these choices we see that the future
pointing unit normal to the Cauchy surface at the central vertex $O$ is just $(n^\mu)_o =
(1,0,0,0)^\mu$ while $(n_\mu)_o= (-1,0,0,0)$. We also see that the tetrad $e_a$ has components
$e^\mu{}_a = \delta^\mu{}_a$ in this RNC frame. Note that $e^\mu{}_a e_\mu{}^b =
\delta_a{}^b$, $e^\mu{}_a e^a{}_\nu = \delta^\mu{}_\nu$, $e^\mu{}_1 = n^\mu$ and $e_\mu{}^1 =
-n_\mu$.

% ============================================================================================
\section{Evolving the leg-lengths}

The legs of the lattice are required to be short geodesic segments. Thus it should come as no
surprise that the evolution of the leg-lengths can be obtained from the equations for the
second variation of arc length. In an earlier paper \cite{brewin:2009-04} we showed that, for
sufficiently short legs, these equations can be written as follows
\begin{align}
\DLsqDt &= -2 N K_{\mu\nu}  \Dxij^\mu \Dxij^\nu + \BigO{L^3}
\label{eqn:ADMDLija}\\[10pt]
\DDLsqDt &= 2 N_{|\alpha\beta} \Dxij^\alpha \Dxij^\beta
\label{eqn:ADMDLijb}\\
              & \quad + 2 N\left(  K_{\mu\alpha}K^\mu{}_\beta 
                                 - R_{\mu\alpha\nu\beta} n^\mu n^\nu \right)
                      \Dxij^\alpha \Dxij^\beta + \BigO{L^3}\notag
\end{align}
For numerical purposes it is somewhat easier to rewrite these in the following form
\begin{align}
\DLsqDt &= -2 N \Pij\label{eqn:ADMLijc}\\[10pt]
\DPijDt &= - N_{|\alpha\beta} \Dxij^\alpha \Dxij^\beta
\label{eqn:ADMDLijd}\\
              & \quad - N\left(  K_{\mu\alpha}K^\mu{}_\beta 
                               - R_{\mu\alpha\nu\beta} n^\mu n^\nu \right)
                      \Dxij^\alpha \Dxij^\beta\notag
\end{align}
in which we have introduced the new variables $\Pij$, one per leg. The $\Kmn$ can be obtained
by a suitable weighted sum of equation (\ref{eqn:ADMDLija}) as described in section
(\ref{sec:Extrinsic}). We have also dropped the truncation terms as these are not used during
a numerical integration.

Clearly, the evolution of the leg lengths requires a knowledge of the Riemann curvatures and to
that end we now present the evolution equations for those curvatures.

% ============================================================================================
\section{Evolving the Riemann curvatures. Pt. 1}
\label{sec:EvolvRiemPt1}

We know that there are only 20 algebraically independent Riemann curvatures in 4 dimensions.
So which should we choose? By a careful inspection of the algebraic symmetries of $\Rmanb$ we
settled upon the following
\begin{gather}
\Rxyxy,\> \Rxyxz,\> \Rxyyz,\> \Rxzxz,\> \Rxzyz,\> \Ryzyz\notag\\
\Rtxxy,\> \Rtxxz,\> \Rtyxy,\> \Rtyxz,\> \Rtyyz,\> \Rtzxy,\> \Rtzyz,\> \Rtzyz\\
\Rtxtx,\> \Rtyty,\> \Rtztz,\> \Rtxty,\> \Rtxtz,\> \Rtytz\notag
\end{gather}

% --------------------------------------------------------------------------------------------
\subsection{Bianchi identities}

Our aim is to use the Bianchi identities to obtain evolution equations for the Riemann
curvatures. We begin by writing down the Bianchi identities at the central vertex, where the
connection vanishes,
\begin{equation}
0 = \dRmanbg + \dRmabgn + \dRmagnb
\label{eqn:BianchiA}
\end{equation}
along with a contracted version of the same equation
\begin{equation}
0 = g^{\mu\gamma} \dRmanbg - \dRabn + \dRanb
\label{eqn:BianchiB}
\end{equation}
This pair of equations, along with the vacuum Einstein field equations, and a judicious choice
of indices will provide us with all of the required evolution equations. This leads to the
following 14 differential equations
\bgroup
\def\v{\noalign{\vskip3pt\vskip 0pt plus 10pt\penalty-250\vskip 0pt plus-10pt}}
\begin{align}
0 &= \dRxyxyt - \dRtyxyx + \dRtxxyy\label{eqn:RiemEvolA}\\\v
0 &= \dRxyxzt - \dRtzxyx + \dRtxxyz\label{eqn:RiemEvolB}\\\v
0 &= \dRxyyzt - \dRtzxyy + \dRtyxyz\label{eqn:RiemEvolC}\\\v
0 &= \dRxzxzt - \dRtzxzx + \dRtxxzz\label{eqn:RiemEvolD}\\\v
0 &= \dRxzyzt - \dRtzxzy + \dRtyxzz\label{eqn:RiemEvolE}\\\v
0 &= \dRyzyzt - \dRtzyzy + \dRtyyzz\label{eqn:RiemEvolF}\\\v
0 &= \dRtyxyt - \dRxyxyx + \dRxyyzz\label{eqn:RiemEvolG}\\\v
0 &= \dRtxxyt + \dRxyxyy + \dRxyxzz\label{eqn:RiemEvolH}\\\v
0 &= \dRtzxyt - \dRxyxzx - \dRxyyzy\label{eqn:RiemEvolI}\\\v
0 &= \dRtzxzt - \dRxzxzx - \dRxzyzy\label{eqn:RiemEvolJ}\\\v
0 &= \dRtxxzt + \dRxyxzy + \dRxzxzz\label{eqn:RiemEvolK}\\\v
0 &= \dRtyxzt - \dRxyxzx + \dRxzyzz\label{eqn:RiemEvolL}\\\v
0 &= \dRtzyzt - \dRxzyzx - \dRyzyzy\label{eqn:RiemEvolM}\\\v
0 &= \dRtyyzt - \dRxyyzx + \dRyzyzz\label{eqn:RiemEvolN}
\end{align}
\egroup

There are of course 20 independent $\Rmanb$, 14 of which are subject to the above evolution
equations while the remaining 6 can be obtained from the vacuum Einstein equations
\begin{align}
0 &= \Rxx = -\Rtxtx + \Rxyxy + \Rxzxz\label{eqn:RiemEvolO}\\
0 &= \Ryy = -\Rtyty + \Rxyxy + \Ryzyz\label{eqn:RiemEvolP}\\
0 &= \Rzz = -\Rtztz + \Rxzxz + \Ryzyz\label{eqn:RiemEvolQ}\\
0 &= \Rxy = -\Rtxty + \Rxzyz\label{eqn:RiemEvolR}\\
0 &= \Rxz = -\Rtxtz - \Rxyyz\label{eqn:RiemEvolS}\\
0 &= \Ryz = -\Rtytz + \Rxyxz\label{eqn:RiemEvolT}
\end{align}
Though these are not differential equations they do, none the less, provide a means to evolve
the 6 curvatures $\Rtxtx,\Rtxty\cdots\Rtytz$.

The important point to note about this system of equations is that it is closed, there are 20
evolution equations for 20 curvatures. The source terms, such as $\dRxyxyx$, could be computed
by importing data from the neighbouring cells, by an appropriate combination of rotations and
boosts, and using a suitable finite difference approximation (see section
(\ref{sec:SourceTerms}) for more details)). In this way the lattice serves as a scaffold on
which source terms such as these can be computed.

% --------------------------------------------------------------------------------------------
\subsection{Constraints}

In deriving the 20 evolution equations of the previous section we used only 6 of the
10 vacuum Einstein equations. Thus the 4 remaining vacuum Einstein equations must be viewed
as constraints. These equations are
\begin{align}
0 &= \Rtt = \phantom{-} \Rtxtx + \Rtyty + \Rtztz\label{eqn:EinsConstA}\\
0 &= \Rtx = \phantom{-} \Rtyxy + \Rtzxz\label{eqn:EinsConstB}\\
0 &= \Rty = - \Rtxxy + \Rtzyz\label{eqn:EinsConstC}\\
0 &= \Rtz = - \Rtxxz - \Rtyyz\label{eqn:EinsConstD}
\end{align}

Finally, we have the following 6 constraints that arise from the Bianchi identities.
\begin{align}
0 &= \dRxyxyz + \dRxyyzx - \dRxyxzy\label{eqn:BianConstA}\\
0 &= \dRxyxzz + \dRxzyzx - \dRxzxzy\label{eqn:BianConstB}\\
0 &= \dRxyyzz + \dRyzyzx - \dRxzyzy\label{eqn:BianConstC}\\
0 &= \dRtxxyz + \dRtxyzx - \dRtxxzy\label{eqn:BianConstD}\\
0 &= \dRtyxyz + \dRtyyzx - \dRtyxzy\label{eqn:BianConstE}\\
0 &= \dRtzxyz + \dRtzyzx - \dRtzxzy\label{eqn:BianConstF}
\end{align}

So all up we have 
20 evolution equations assembled from the %
   14 differential equations (\ref{eqn:RiemEvolA}--\ref{eqn:RiemEvolN}) and %
    6 algebraic equations (\ref{eqn:RiemEvolO}--\ref{eqn:RiemEvolT}) plus 
10 constraints comprising 
    4 Einstein equations (\ref{eqn:EinsConstA}--\ref{eqn:EinsConstD}) and 
    6 Bianchi identities (\ref{eqn:BianConstA}--\ref{eqn:BianConstF}). 
This is a such a simple system that it allows simple questions to be explored and answered
with ease. The questions that we will address are
\begin{enumerate}
\item Are the constraints preserved by the evolution equations?
\item Do the evolution equations constitute a hyperbolic system? 
\end{enumerate}
For both questions the answer is yes and we shall now demonstrate that this is so.

% --------------------------------------------------------------------------------------------
\subsection{Constraint preservation}

In the following discussion we will assume that, by some means, we have constructed an initial
data set for the 20 $\Rmanb$. That is, the 20 $\Rmanb$ are chosen so that the 10 constraints
(\ref{eqn:EinsConstA}--\ref{eqn:BianConstF}) vanish at the central vertex of \emph{every}
computational cell in the lattice.

We will also need the trivial result that
\begin{equation}
R = 2\left(\Rxyxy + \Rxzxz + \Ryzyz\right)
\label{eqn:ScalarR}
\end{equation}
which follows directly from equations (\ref{eqn:RiemEvolO},\ref{eqn:RiemEvolP},%
\ref{eqn:RiemEvolQ},\ref{eqn:EinsConstA}).

Consider now the constraint $0=\Rtz$. By assumption, this constraint is satisfied on the
initial slice. To demonstrate that it continues to hold throughout the evolution we need to
show that $0=\dRtzt$. From (\ref{eqn:EinsConstD}) this requires us to show that
$0=\dRtxxzt+\dRtyyzt$. Using (\ref{eqn:RiemEvolK},\ref{eqn:RiemEvolN}) we see that
\begin{equation*}
\dRtxxzt+\dRtyyzt = -\dRxyxzy-\dRxzxzz+\dRxyyzx-\dRyzyzz
\end{equation*}
however on the initial slice we also have, by assumption, (\ref{eqn:BianConstA})
\begin{equation*}
0 = \dRxyxyz+\dRxyyzx-\dRxyxzy
\end{equation*}
which when combined with the previous equation leads to
\begin{equation*}
\dRtxxzt+\dRtyyzt = -\left(\Rxyxy+\Rxzxz+\Ryzyz\right)_{,z}
\end{equation*}
But by equation (\ref{eqn:ScalarR}) we see that the right hand side is just $-\dRz/2$ and as
$R=0$ across the initial slice we also have that $0=\dRz$ at every central vertex. This
completes the proof. The two other constraints, $0=\Rty$ and $0=\Rtx$, can be dealt with in a
similar fashion.

All that remains is to show that $0=\Rtt$ is conserved. We proceed in a manner similar to the
above. First we use $\Rtt=\Rxyxy + \Rxzxz + \Ryzyz$ and then use equations 
(\ref{eqn:RiemEvolA},\ref{eqn:RiemEvolD},\ref{eqn:RiemEvolF}) to compute
the time derivative
\begin{align*}
\left(\Rxyxy + \Rxzxz + \Ryzyz\right)_{,t}
 &= \dRxyxyt + \dRxzxzt + \dRyzyzt\\
 & =\phantom{+} \dRtyxyx-\dRtxxyy\\
 &\>\phantom{=}+\dRtzxzx-\dRtxxzz\\
 &\>\phantom{=}+\dRtzyzy-\dRtyyzz\\
 &= \dRtxx+\dRtyy+\dRtzz
\end{align*}
where the last line arose by inspection of equations
(\ref{eqn:EinsConstB}--\ref{eqn:EinsConstD}). But $0=\Rmn$ at every central vertex on the
initial slice. Thus $0=\dRmni$, $i=x,y,z$ on the central vertex which in turn shows that
$0=\dRttt$ on the initial slice.

A key element in the above proofs was the use of constraints based on the Bianchi
identities. The question now must be -- do the evolution equations preserve those constraints?
The answer is yes which we will now demonstrate on a typical case. Consider the constraint
(\ref{eqn:BianConstA})
\begin{equation*}
0 = \dRxyxyz + \dRxyyzx - \dRxyxzy 
\end{equation*}
We know this to be true on the initial slice and we need to show that the evolution equations
(\ref{eqn:RiemEvolA}--\ref{eqn:RiemEvolN}) guarantee that it will be satisfied on all
subsequent slices. The calculations follow a now familiar pattern,
\begin{align*}
\left(\dRxyxyz + \dRxyyzx - \dRxyxzy\right){}_{,t}
   &= \dRxyxytz + \dRxyyztx - \dRxyxzty\\
   &=\phantom{+}\left( \dRtyxyx-\dRtxxyy\right){}_{,z}\\
   &\phantom{=}{}\>+\left(\dRtzxyy-\dRtyxyz\right){}_{,x}\\
   &\phantom{=}{}\>-\left(\dRtzxyx-\dRtxxyz\right){}_{,y}\\
   &=0
\end{align*}
The same analysis can be applied to the remaining constraint equations.

% --------------------------------------------------------------------------------------------
\subsection{Hyperbolicity}

Our approach to proving hyperbolicity will be quite simple. We will manipulate the evolution
equations (\ref{eqn:RiemEvolA}--\ref{eqn:RiemEvolN}) to demonstrate that each of our 20
$\Rmanb$ satisfies the standard second order wave equation.

Let us start with a simple example, equation (\ref{eqn:RiemEvolA}). We take one further time
derivative, commute the mixed partial derivatives and then use equations
(\ref{eqn:RiemEvolG},\ref{eqn:RiemEvolH}) to eliminate the single time derivative. This leads
to
\begin{equation*}
0 = \dRxyxytt-\dRxyxyxx-\dRxyxyyy+\dRxyyzzx-\dRxyxzzy
\end{equation*}
However, we also have $0=\dRxyxyz+\dRxyyzx-\dRxyxzy$, which allows us to reduce the last two
terms of the previous equation to just $-\dRxyxyzz$. Thus we have
\begin{equation*}
0 = \dRxyxytt-\dRxyxyxx-\dRxyxyyy-\dRxyxyzz
\end{equation*}
This is the standard flat space wave equation for $\Rxyxy$. A similar analysis shows that
$\Rxzxz$, $\Ryzyz$, $\Rxyxz$, $\Rxyyz$ and $\Rxzyz$ are also solutions of the wave equation.

We now turn to the 8 $\Rmanb$ in which the indices $\mu\alpha\nu\beta$ contain just one $t$.
The proof (that each such $\Rmanb$ satisfies the wave equation) differs from the above only
in the way the Bianchi identities are used. Applying the first few steps outlined above to
equation (\ref{eqn:RiemEvolG}) leads to
\begin{align*}
0 = \dRtyxytt&-\dRtyxyxx-\dRtyxyyy-\dRtyxyzz\\
             &+\dRtyxyyy+\dRtxxyxy+\dRtzxyzy
\end{align*}
in which we have deliberately introduced the pair of terms $\dRtyxyyy$ to aid in the following
exposition. The last three terms can be dealt with as follows. First notice that
\begin{align*}
\dRtyxyyy+\dRtxxyxy+\dRtzxyzy 
   &= \left( \dRtyxyy+\dRtxxyx+\dRtzxyz \right){}_{,y}\\
   &= \left( -\R^{\mu}{}_{txy,\mu} \right){}_{,y}\\
   &= \left( -\dRtyx+\dRtxy\right){}_{,y}
\end{align*}
where in last line we have used the contracted Bianchi identity 
$0=\R^{\mu}{}_{\nu\alpha\beta,\mu}-\R_{\nu\beta,\alpha}+\R_{\nu\alpha,\beta}$. But we know
that $0=\Rmn$ at every central vertex, thus all of its partial derivatives will be zero
and so the each term on the right hand vanishes leading to our desired result
\begin{equation*}
0 = \dRtyxytt-\dRtyxyxx-\dRtyxyyy-\dRtyxyzz
\end{equation*}

Finally we note that the remaining 6 $\Rmanb$, that is those that carry two $t$'s in their
indices, are linear combinations of the previous 14 $\Rmanb$, see equations
(\ref{eqn:RiemEvolO}--\ref{eqn:RiemEvolT}), and thus will also be solutions of the wave
equation. Thus we have shown, as claimed, that all 20 $\Rmanb$ satisfy the wave equation.

% ============================================================================================
\section{Evolving the Riemann curvatures. Pt. 2}
\label{sec:EvolvRiemPt2}

There are two problems in the forgoing analysis. The first problem is that we chose a unit
lapse function when presenting the evolution equations
(\ref{eqn:RiemEvolA}--\ref{eqn:RiemEvolN}). We can easily remedy this problem by making a
simple vertex dependent coordinate substitution $t=Nt'$ in each of the evolution equations.

The second problem is somewhat more of a challenge. It stems from the simple fact that each
computational cell is local in both space and time and therefore no single RNC can be used to
track the evolution for an extended period of time. We will have no choice but to jump
periodically to a new RNC frame. But how might we do this? One approach goes as follows. Build,
on the world line of a typical vertex, a pair of distinct but overlapping cells, with one cell
lying slightly to the future of the other. Then evolve the curvatures in the frame of one cell
into the overlap region followed by a coordinate transformation to import the newly evolved
curvatures into the frame of the future cell. This completes one time step of the integration
whereupon the whole process can be repeated any number of times along the vertex world line. A
useful improvement on this is to use a local tetrad to construct scalars thus avoiding the need
for explicit coordinate transformations when passing from one cell to the next. The price we
pay for this is that we have to account for the evolution of the tetrad along the world line.
As we shall see this is rather easy to do (essentially we project the tetrad onto the legs of
the lattice). We will explore this method first on a simple example before presenting the
computations for the curvature evolution equations.

% --------------------------------------------------------------------------------------------
\subsection{A simple example}

In this example we will suppose that we have a vector $W^\mu$ that evolves along the
world line of the central vertex according to
\begin{equation}
\frac{dW^\mu}{dt'} = NF^\mu\label{eqn:SimpleDot}
\end{equation}
Our aim is to obtain a related equation that describes the evolution of the vector along the
whole length of the world line, not just the short section contained within this one cell.

\begin{figure}[t]
\centerline{\includegraphics[width=\textwidth]{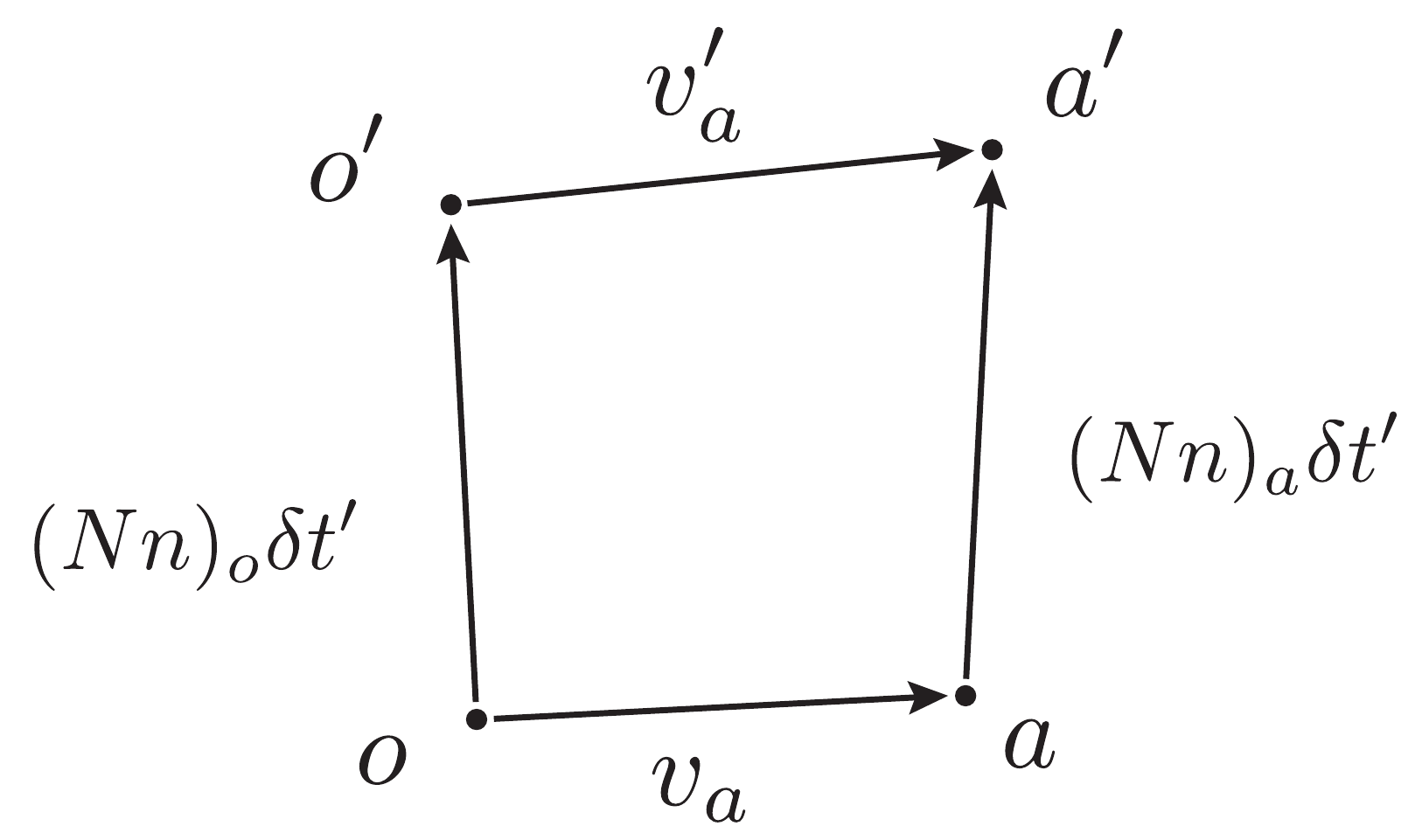}}
\vskip0.5cm
\caption{\normalfont Here we show the evolution of one leg $(oa)$ within one computational
cell. Clearly the four vectors form a closed loop and thus $(Nn)_o\delta t'+v'_a= v_a +
(Nn)_a) \delta t'$ which leads directly to equation (\ref{eqn:VNcommute}).}
\label{img:Vectors}
\end{figure}

Suppose that we have an orthonormal tetrad $e_a=\ema\partial_\mu$, $a=1,2,3,4$ on $\omega$
with $e_1$ aligned to $n^\mu\partial_\mu$, the future pointing normal to $\omega$, and that we
have aligned the RNC coordinate axes with the tetrad (note how this gives precedence to the
tetrad over the coordinates). Thus at the central vertex of $\Omega$ we have
\begin{gather*}
\ea = \partial_a\>,\quad
\ema = \delta^\mu{}_a\>,\quad
\Ema = \delta_\mu{}^a\\[3pt]
 n^\mu = e^\mu{}_1 \>,\quad
-n_\mu = e_\mu{}^1 \\[3pt]
e^\mu{}_a e_\mu{}^b = \delta_a{}^b\>,\quad
e^\mu{}_a e_\nu{}^a = \delta_\mu{}^\nu\\[3pt]
\gmn = \diag(-1,1,1,1)\>,\quad
\gab = \diag(-1,1,1,1)
\end{gather*}
We now propose the following evolution equations along the world line of the central vertex in
$\Omega$.
\begin{gather}
\frac{de^\mu{}_1}{dt'} = e^\mu{}_i \nabla^i N
\>,
\quad
\frac{de_\mu{}^1}{dt'} = -e_\mu{}^i \nabla_i N
\label{eqn:TetradDtA}\\[3pt]
\frac{de^\mu{}_i}{dt'} = e^\mu{}_1 \nabla_i N
\>,
\quad
\frac{de_\mu{}^i}{dt'} = -e_\mu{}^1 \nabla^i N
\>,
\quad i=2,3,4
\label{eqn:TetradDtB}
\end{gather}
where $\nabla_i N = (\bot N_{,\nu}) e^\nu{}_i$ and $\nabla^i N = (\bot N^{,\nu}) e_\nu{}^i$,
$i=2,3,4$. What can we say about the evolved data? First, note that the orthonormal conditions
are preserved, that is
\begin{equation*}
\frac{de^\mu{}_a e_\mu{}^b}{dt'} = 0\>,\quad
\frac{de^\mu{}_a e_\nu{}^a}{dt'} =0
\end{equation*}
Thus the tetrad obtained by integrating the above equations will remain orthonormal
along the world line of the central vertex. Second, using
\begin{equation}
\left(N n^\mu\right)_{;\nu} = N_{,\nu} n^\mu - \bot(N^{,\mu}) n_\nu - N K^\mu{}_{\nu}
\label{eqn:DerivNormal}
\end{equation}
to compute $dn^\mu/dt' = n^\mu{}_{;\nu}\left(Nn^\nu\right)$ we see that
\begin{equation}
\frac{de^\mu{}_1}{dt'} =  \frac{dn^\mu}{dt'}\>,\quad
\frac{de_\mu{}^i}{dt'} = -\frac{dn_\mu}{dt'}
\end{equation}
which shows that $e^\mu{}_1 = n^\mu$ and $e_\mu{}^1 = -n_\mu$ everywhere along the world line.
That is, $e^\mu{}_1$ remains tied to the world line. All that remains is to account for how
the tetrad rotates around the world line. This we shall do by evolving the projections of the
$e^\mu{}_i$, $i=2,3,4$ onto the legs of the lattice. Let $v_a=v^\mu{}_a\partial_\mu$,
$a=1,2,3$ be any three distinct legs of the lattice attached to the central vertex. Now
consider a short time step in which the vector $v_a$ sweeps out a short quadrilateral in
spacetime (see figure (\ref{img:Vectors})). The upper and lower edges will be the past and
future versions of $v_a$ while the remaining two sides will be generated by the word lines of
the vertices that define $v_a$. Since we have assumed at the outset that all vertices evolve
normal to the Cauchy surface we see that these vertical vectors correspond to $Nn^\mu$. The
important point is that this set of four vectors forms a closed loop, in short the vectors
$v_a$ and $Nn^\mu\partial_\mu$ commute, thus
\begin{equation}
v^\mu{}_{a;\nu}\left(Nn^\nu\right) = v^\nu{}_a\left(Nn^\mu\right)_{;\nu}
\label{eqn:VNcommute}
\end{equation}
The left hand side is simply $dv^\mu/dt'$, while the right hand side can be expanded using 
(\ref{eqn:DerivNormal}). This leads to
\begin{equation}
\frac{dv^\mu{}_a}{dt'} = \left(  N_{,\nu} n^\mu - N K^\mu{}_\nu \right) v^\nu{}_a
\label{eqn:VaDt}
\end{equation}
where we have dropped the term involving $n_\mu v^\mu{}_a$ as this would be
$\BigO{L^m}$ with $m\ge2$ while the remaining terms are all $\BigO{L}$.

We are now ready to construct our scalar evolution equations. Let $\W_a :=W_\mu\ema$ and
$\v_a{}^b := v^\mu{}_a e_\mu{}^b$ then
\begin{gather*}
\frac{d\W_a}{dt} = \frac{dW_\mu}{dt'}\ema + W_\mu \frac{d\ema}{dt'}\\[5pt]
\frac{d\v_a{}^b}{dt'} = \frac{dv^\mu{}_a}{dt'}e_\mu{}^b + v^\mu{}_a \frac{de_\mu{}^b}{dt'}\>,
\quad b=1,2,3,4
\end{gather*}
Each of these equations can be re-cast entirely in terms of the scalars by first using
(\ref{eqn:TetradDtA},\ref{eqn:TetradDtB},\ref{eqn:VaDt}) to eliminate the time derivatives on
the right hand side followed by the substitutions $W_\mu=\W_a e_\mu{}^a$ and $v^\mu{}_a =
\v_a{}^b e^\mu{}_b$. This leads to
\begin{gather}
\frac{d\W_n}{dt'} = N \F_n + \W_i \nabla^i N\\[5pt]
\frac{d\W_i}{dt'} = N \F_i + \W_n \nabla_i N\>,\quad i=2,3,4\\[5pt]
\frac{d\v_a{}^1}{dt'} = \frac{1}{N}\frac{dN}{dt'}\v_a{}^1\\[5pt]
\frac{d\v_a{}^i}{dt'} = - N \K^i{}_j \v_a{}^j\>,\quad i,j=2,3,4\>,\>\>a=1,2,3\label{eqn:EdotV}
\end{gather}
where we have introduced the scalars $\W_n = W_\mu n^\mu$, $\F_n = F_\mu n^\mu$, $\F_i = F_\mu
e^\mu{}_i$, and $\K^i{}_j = K^\mu{}_\nu e_\mu{}^i e^\nu{}_j$. These are our final equations.
They are valid along the whole length of the world line, not just the part contained in one
cell.

Equation (\ref{eqn:EdotV}) describes the motion of the tetrad relative to the legs of the
lattice. As we integrate forward in time we can use the values of $\v_a{}^i$ to locate the
tetrad within the computational cell. If we chose to construct an RNC within the cell then we
can go one step further and recover the values of $e^\mu{}_i$ and the $W^\mu$.

% --------------------------------------------------------------------------------------------
\subsection{Curvature evolution equations}
\label{sec:NewRiemEvolve}

Now we can return to the task of constructing the generalised evolution equations for the
curvatures. We start by introducing a pair of relations between the tetrad and coordinate
components of the curvature tensor
\begin{align*}
\rabcd &= \Rmanb \ema \eab \enc \ebd\\[3pt]
\Rmanb &= \rabcd \Ema \Eab \Enc \Ebd
\end{align*}
and then forming a typical evolution equation
\begin{equation}
\DrabcdDt = \DRmanbDt\ema\eab\enc\ebd + \Rmanb \frac{d\left(\ema\eab\enc\ebd\right)}{dt'}
\end{equation}
with each $d/dt'$ term on the right hand side replaced by a suitable combination of the
existing evolution equations, (\ref{eqn:RiemEvolA}--\ref{eqn:RiemEvolN}) for the curvature
terms and (\ref{eqn:TetradDtA},\ref{eqn:TetradDtB}) for the tetrad terms.

Rather than working through all 14 equations we will demonstrate the procedure on just one
equation (\ref{eqn:RiemEvolA}) leaving the remaining equations (but not their working) to the
Appendix. So our starting point is
\begin{equation*}
\DrxyxyDt = \frac{d\Rxyxy}{dt'} + \Rmanb\frac{d\left(\emx\eay\enx\eby\right)}{dt'}
\end{equation*}
and using (\ref{eqn:RiemEvolA}) we obtain
\begin{equation*}
\DrxyxyDt = \dRtyxyx - \dRtxxyy + \Rmanb\frac{d\left(\emx\eay\enx\eby\right)}{dt'}
\end{equation*}
Finally we use (\ref{eqn:TetradDtA},\ref{eqn:TetradDtB}) to eliminate the time derivative of
$\ema$, leading to
\begin{align}
\DrxyxyDt &= \dRtyxyx - \dRtxxyy\notag\\
          &\quad + \rtyxy\dNx - \rtxxy\dNy + \rtyxy\dNx - \rtxxy\dNy
          \label{eqn:NewEvolve}
\end{align}
This is as far as we need go, though it is tempting to make the substitutions
$\Rtyxy=\rabcd\Eta\Eyb\Exc\Eyd$ and $\Rtxxy=\rabcd\Eta\Exb\Exc\Eyd$. But that is not really
necessary as we can defer those substitutions until we actually need values for the stated
partial derivatives. This is described in more detail in section (\ref{sec:SourceTerms}).

Note that when introducing the lapse function by the substitution $t=Nt'$ we have not made
explicit the coordinate transformation on the curvatures other than to use distinct labels $t$
and $t'$. In this way we use $t'$ as an integration parameter on the world line of each vertex
while retaining the original coordinates $(t,x,y,z)$ as the local Riemann normal coordinates
(and thus at any point on the world line we continue to have $(\gmn)_o=\diag(-1,1,1,1)$). We
choose to maintain this distinction between $t$ and $t'$ not only to keep the equations tidy
but also because it leaves the equations in a simple form well suited to numerical
integrations. 

Clearly the above procedure can be applied directly to each of the remaining 13 curvature
evolution equations. The final results for all 14 equations can be found in the Appendix.

% --------------------------------------------------------------------------------------------
\subsection{Hyperbolicity and constraint preservation}
\label{sec:CheckHyperbol}

It is natural to ask if the new system of evolution equations are
hyperbolic and also, are the new constraints preserved by the new evolution equations?
The answer to both questions is yes and we will demonstrate this as follows. 

Given that $\rabcd = \Rmanb \ema \eab \enc \ebd$ we see that
\begin{equation*}
\DrabcdDef = \DRmanbDrt \ema \eab \enc \ebd \ere \etf
           + {\cal V}_{abcdef}
           \left(R,N,\partial R,\partial N,\partial^2 N\right)
\end{equation*}
where ${\cal V}_{abcdef}$ is a function of $\Rmanb$, $N$ and the indicated partial
derivatives. Importantly, ${\cal V}_{abcdef}$ does not contain any second partial derivatives
of the curvatures. We have previously shown that, at the central vertex, each $\Rmanb$
satisfies a wave equation of the form $0=g^{\rho\tau}\DRmanbDrt$ with
$g^{\rho\tau}=\diag(-1,1,1,1)$. Thus we find that
\begin{equation*}
g^{ef}\DrabcdDef 
   = g^{ef} {\cal V}_{abcdef}
     \left(R,\partial R,N,\partial N,\partial^2 N\right)
\end{equation*}
where $g^{ef} = \diag(-1,1,1,1)$. It follows that each
$\rabcd$ satisfies a wave equation with source terms and therefore we have shown that the new
evolution equations constitute a hyperbolic system.

A similar analysis can be applied to the constraints. We begin by writing a typical
differential constraint (\ref{eqn:BianConstA}--\ref{eqn:BianConstF}) in the form
\begin{equation*}
0 = \Wmanb(\partial R)
\end{equation*}
where the right hand side depends only on the the first derivatives of $\Rmanb$. Introducing
the lapse function is trivial (there are no time derivatives, so the equation is unchanged).
If we define the frame components $\Wabcd$ by
\begin{equation*}
\Wabcd = \Wmanb \ema \eab \enc \ebd
\end{equation*}
then we find
\begin{equation*}
\dWabcd = \dWmanbr \ema \eab \enc \ebd \ert
        + \Wmanb \left( \ema \eab \enc \ebd \right)_{,t}
\end{equation*}
and as we have previously shown that $\Wmanb=0$ and $\dWmanbr=0$ it follows that $\Wabcd=0$
and $\dWabcd=0$. It is easy to see that the same procedure can be applied to the remaining
constraints (\ref{eqn:RiemEvolO}--\ref{eqn:EinsConstA}) with the same outcome. Thus we have
shown that the new constraints are conserved by the new evolution equations.

% ============================================================================================
\section{Coordinates}
\label{sec:Coords}

There are at least two instances where the vertex coordinates are required. First, when
constructing the transformation matrix used when importing data from neighbouring cells.
Second, as part of the time integration of leg-lengths, equations
(\ref{eqn:ADMDLija}--\ref{eqn:ADMDLijb}). They are also required when computing the
extrinsic curvatures (\ref{sec:Extrinsic}) and the hessian (\ref{sec:Hessian}).

Recall that within each cell we employ two distinct coordinate frames, one is tied to the
tetrad associated with the central vertex while the other is aligned with the lattice. Both
frames share the central vertex as the origin. We will describe first how to construct the
lattice coordinates, which we will denote by $y^\mu$, followed by the tetrad coordinates,
denoted by $x^\mu$. The lattice coordinates are only ever used in the construction of the
tetrad coordinates, once these are known then the lattice coordinates can be discarded. Note
that terms such as $R_{xyxy}$, $K_{xy,z}$ etc. are referred to the tetrad coordinates.

For a large part of this discussion we will be concerned mainly with the scaling of the
coordinates with respect to the typical lattice scale (\eg to establish that $t=\BigO{L^2}$).
This applies equally well to both coordinate frames and so, to be specific, we will present
the arguments in terms of the tetrad coordinates. Once we have sorted out these scaling issues
we will compute the lattice coordinates followed by the tetrad coordinates.

Our first task will be to construct the piece of the Cauchy surface that is covered by a
typical computational cell. Recall that we view the Cauchy surface to be a smooth
3-dimensional surface that passes through each vertex of the lattice and that it shares with
the lattice, at each vertex, the same future pointing unit normal and second fundamental form
(the extrinsic curvatures). In our local Riemann normal coordinates we wish to construct an
equation of the form $0=-t+f(x^u)$ that passes through the vertices of this computational cell
and with given extrinsic curvature at the central vertex. For this we use the familiar
definition that $\delta n^\mu = -K^\mu{}_{\nu}\delta x^\nu$ for the small change in the unit
normal under a displacement across the Cauchy surface. If we take the displacement to be from
the central vertex $(o)$ to a nearby vertex $(a)$ then we have
\begin{equation}
n^\mu_a - n^\mu_o = -K^\mu{}_{\nu} x^\nu_{a}
\end{equation}
But we chose the coordinates so that $n^\mu_o = (1,0,0,0)^\mu$ while for the surface
$0=-t+f(x^u)$ the unit normal at $(a)$ is simply $n^\mu_a = g^{\mu\nu} (-1,f_{,u})_{\nu}/M =
(1,f_{,u})^\mu/M$ where $M=1+\BigO{L^2}$ is a normalization factor. Thus we have
$(1,f_{,u})^\mu = (1,0,0,0)^\mu - K^\mu{}_{\nu} x^\nu_{a} + \BigO{L^2}$ and this is easily
integrated to give
\begin{equation}
t_a = -\frac{1}{2}\Kmn x^\mu_a x^\nu_a + \BigO{L^3}
\label{eqn:TimeCoord}
\end{equation}
Note that since $K^\mu{}_\nu n^\nu = 0$ we can use this last equation to compute the time
coordinates for each vertex in the computational cell (given the spatial coordinates $x^u_a$
and the extrinsic curvatures $\Kuv$).

Consider the geodesic segment that joins the central vertex $(o)$ to a typical nearby vertex
$(a)$. Then from the definition of Riemann normal coordinates we have
\begin{equation}
x^\mu_a = m^\mu_a L_{oa}
\end{equation}
where $m^\mu_a$ is the unit tangent vector to the geodesic at $(o)$.\footnote{Actually, by
virtue of the fact that the path is a geodesic segment expressed in Riemann normal
coordinates, the values for $m^\mu_a$ are constant along the geodesic.} Thus it follows that
\begin{equation}
\vert x^\mu_a\vert = \BigO{L}
\end{equation}
for each vertex in the computational cell. Combining this with the above equation
(\ref{eqn:TimeCoord}) for $t_a$ shows that
\begin{equation}
\vert t_a\vert = \BigO{L^2}
\end{equation}
This result could also be inferred from the simple observation that $m^t\rightarrow0$ as
$L\rightarrow0$ (this is a consequence of the smoothness of the Cauchy surface at $(o)$).

We turn now to the simple question -- How accurate do we need the coordinates to be? That is,
if ${\tilde x}^\mu_i$ are the exact Riemann normal coordinates for vertex $i$, then how large
can we allow $\vert x^\mu_i - {\tilde x}^\mu_i\vert$ to be? The answer can be found by a
simple inspection of the evolution equations (\ref{eqn:ADMDLija}--\ref{eqn:ADMDLijb}). The
truncation terms in those equations are $\BigO{L^3}$ thus we can safely get by with
$\BigO{L^2}$ errors in the coordinates, that is
\begin{equation}
\vert x^\mu_i - {\tilde x}^\mu_i\vert = \BigO{L^2}
\end{equation}

The good news is that such coordinates are readily available -- flat space will do. 
To see that this is so, assume, for the moment, that we have estimates for the $\Kmn$ and then
look back at equations (\ref{eqn:TimeCoord},\ref{eqn:RNCLsq}). This is a coupled system of
equations for the coordinates $(t,x,y,z)^\mu_a$ for each vertex in the computational cell. We
are fortunate to have an explicit equation for the time coordinates, namely
(\ref{eqn:TimeCoord}). This allows us, in principle, to eliminate each time coordinate that
appears in equation (\ref{eqn:RNCLsq}). The result would be a set of equations for the spatial
coordinates $x^u_a$. In the following we will not make this elimination explicit but take it
as understood that such a process has been applied. We will have a little more to say on this
matter in a short while.

For a typical vertex $(l)$ we will need to compute three spatial coordinates and thus we look
to the legs of a tetrahedron. Suppose that that tetrahedron has vertices $(ijkl)$ and suppose
that we have computed, by some means, the exact Riemann normal coordinates $\Tx^\mu$ for
vertices $(ijk)$. The exact coordinates ${\tilde x}^\mu_l$ for vertex $(l)$ could be obtained
by solving the system of equations
\begin{equation}
\Lsq_{al} = \gmn (\Tx^\mu_a-\Tx^\mu_l)(\Tx^\nu_a-\Tx^\nu_l)
          - \frac{1}{3} \Rmanb \Tx^\mu_a \Tx^\nu_a \Tx^\alpha_l \Tx^\beta_l
          \quad\quad a=i,j,k
\label{eqn:LSQCurved}
\end{equation}
but we could also construct flat space coordinates $x^\mu_l$ for vertex $l$ by solving the
system
\begin{equation}
\Lsq_{al} = \gmn (\Tx^\mu_a-x^\mu_l)(\Tx^\nu_a-x^\nu_l)
            \quad\quad a=i,j,k
\label{eqn:LSQFlat}
\end{equation}
From the last equation we conclude that $\vert \Tx^\mu_a-x^\mu_l\vert=\BigO{L}$ for $a\not=l$.
Next, make the trivial substitution $\Tx^\mu_l=x^\mu_l+(\Tx^\mu_l-x^\mu_l)$ in the first term
in (\ref{eqn:LSQCurved}), expand and use (\ref{eqn:LSQFlat}) to obtain
\begin{align*}
 0 = -2 \gmn(\Tx^\mu_a-x^\mu_l)(\Tx^\nu_l-x^\nu_l)
     &+ \gmn(\Tx^\mu_l-x^\mu_l)(\Tx^\nu_l-x^\nu_l)\\
     &- \frac{1}{3} \Rmanb \Tx^\mu_a \Tx^\nu_a \Tx^\alpha_l \Tx^\beta_l
     \quad\quad a=i,j,k
\end{align*}
and as each $\Tx^\mu_a=\BigO{L}$ for $a=i,j,k$ we easily see that
\begin{equation}
\vert \Tx^u_l - x^u_l\vert = \BigO{L^3}
\end{equation}

The fly in the ointment in the above analysis is the assumption that we knew the $\Kmn$ (and
thus we could eliminate the $t_a$). This is not exactly correct for the $\Kmn$ are found by
solving equations (\ref{eqn:ADMDLija}) which in turn requires the coordinates $x^u_a$ which we
have yet to compute (at that stage). Luckily, this is not a major problem. Look carefully at
equation (\ref{eqn:LSQCurved}) and recall that $\gmn = \diag(-1,1,1,1)$. Thus the $t$-terms
will appear only in the form $-(\Tt_a-t_l)^2$ and in the curvature terms of the form
$\R_{tuvw}t_a x^u_a x^v_l x^w_l$. The point to note is that since $t=\BigO{L^2}$ we see that
each of these terms is $\BigO{L^n}$ with $n\ge4$ and thus they have no effect on the above
analysis. Thus even though we argued previously that we should eliminate the $t_a$ using
equation (\ref{eqn:TimeCoord}) the above argument shows that we can put $t_a=0$ without harm.

Our final calculation concerns the errors induced in $t_a$ by using the approximate $x^u_a$
and $\Kmn$ rather than their exact counterparts. Our analysis is very similar to that just
presented. We start with the two sets of equations, the approximate and exact equations,
\begin{equation}
2t_a = -\Kuv x^u_a x^v_a\quad\quad\text{and}\quad\quad
2\Tt_a = -\TKuv \Tx^u_a \Tx^v_a
\end{equation}
We will assume that $\vert\TKuv - \Kuv\vert$ is at least $\BigO{L}$ (this is one assumption
that we will not relax at a later stage). Then we make the trivial substitution
$\Tx^u_l=x^u_l+(\Tx^u_l-x^u_l)$ as above to obtain
\begin{equation}
2\Tt_a = 2t_a -  \left(\TKuv-\Kuv\right)x^u x^v
              - 2\TKuv x^u_a \left(\Tx^v_a-x^v_a\right)
              -  \TKuv \left(\Tx^u_a-x^u_a\right) \left(\Tx^v_a-x^v_a\right)
\end{equation}
Using $x^u_a=\BigO{L}$, $\Tx^u_a=\BigO{L}$ and $\vert\TKuv - \Kuv\vert=\BigO{L}$ we find that
\begin{equation}
\vert\Tt_a - t_a\vert = \BigO{L^3}
\end{equation}

% --------------------------------------------------------------------------------------------
\subsection{The lattice coordinates}
\label{sec:LatticeCoords}

We return now to the concrete question of how to compute the vertex coordinates within one
computational cell. We will first compute the lattice coordinates $y^\mu$ followed by the
tetrad coordinates $x^\mu$. Our present challenge is to find the solutions of the coupled
system of equations
\begin{equation}
\Lsq_{ab} = \gmn (y^\mu_a-y^\mu_b)(y^\nu_a-y^\nu_b)
\end{equation}
for a suitable subset of the legs $(ab)$ in the computational cell (equal in number to the
number of unknown coordinates). The problem here is that if we treat this as a system of
equations for the spacetime coordinates $(t,x,y,z)^\mu_a$ it is extremely unlikely that we
will find any solutions (or if we do then the numerics will almost certainly be extremely
unstable). The reason is quite simple -- the vertices are assumed to lie within one
3-dimensional Cauchy surface. This suggest that we should use the above equations to
determine the spatial coordinates $(x,y,z)^u_a$ with the time coordinates found by other
considerations. Fortunately we already know, from the above analysis, that each $\vert
t_a\vert=\BigO{L^2}$ while $\vert y^u_a\vert = \BigO{L}$. Thus we see that all terms
involving the $t_a$ are $\BigO{L^4}$ and thus will be consumed by the $\BigO{L^4}$ truncation
errors inherent in the above equation (as an approximation to equation
(\ref{eqn:LSQCurved})). So we may safely discard all the of the $t_a$ terms in the above
equations. The next trick that we will use is the observation that the coordinates can be
computed one vertex at a time. This is easily shown by direct construction. Consider a
typical tetrahedron with vertices $(oijk)$ where $(o)$ is the central vertex and suppose we
have computed the coordinates for $(ojk)$. Our task now is to solve the following equations
\begin{align}
\Lsqok &= \phantom{2}\guv y^u_k y^v_k\\
\Lsqok+\Lsqoi-\Lsqik &= 2 \guv y^u_i y^v_k\\
\Lsqok+\Lsqoj-\Lsqjk &= 2 \guv y^u_j y^v_k
\end{align}
where the last pair of equations were obtained by expanding $\Lsqab = \guv
(y^u_a-y^u_b)(y^v_a-y^v_b)$. A simple calculation shows that the solution is given by
\cite{brewin:1998-02}
\begin{equation*}
y^u_k = P y^u_i + Q y^u_j + R n^u
\end{equation*}
where
\begin{gather*}
n^u = g^{uv} \epsilon^{xyz}_{vrs} y^r_i y^s_j\\[5pt]
P=\frac{m_{ik}\Lsqoj-m_{jk}m_{ij}}{\Lsq_n}\quad
Q=\frac{m_{jk}\Lsqoi-m_{ik}m_{ij}}{\Lsq_n}\\[5pt]
R=\pm\frac{\left(\Lsqok-P^2\Lsqoi-Q^2\Lsqoj-2PQm_{ij}\right)^{1/2}}{L_n}\\[5pt]
\Lsq_n = \Lsqoi\Lsqoj-m^2_{ij}
\end{gather*}
and where the $m_{ab}$ are defined by
\begin{gather*}
2m_{ij} = \Lsqoi+\Lsqoj-\Lsqij\\[5pt]
2m_{ik} = \Lsqoi+\Lsqok-\Lsqik\quad\quad\quad
2m_{jk} = \Lsqoj+\Lsqok-\Lsqjk
\end{gather*}
The two solutions, one for each choice of the $\pm$ sign, correspond to the two possible
locations of the third vertex $(k)$, one on each side of the plane containing the triangle
$(oij)$. Which choice is taken will depend on the design of the lattice. A systematic choice
can be made by noting that the vectors $y^u_i$, $y^u_j$ and $n^u$ form a right handed system.
With $R > 0$ the vector $y^u_k$ lives on the same side of the plane as $n^u$.

To complete the picture we need coordinates for the first two vertices $(1)$ and $(2)$.
Since we chose to align our coordinates so that the $x$-axis passed through vertex $(1)$
while the vertex $(2)$ is contained in the $xy$-plane we must have $y^u_1=(A,0,0)^u$ and
$y^u_2=(B,C,0)$ for some numbers $A>0$, $B$ and $C>0$ such that
\begin{align*}
\Lsq_{01} &= \phantom{2}\guv y^u_1 y^v_1\\[3pt]
\Lsq_{02} &= \phantom{2}\guv y^u_2 y^v_2\\[3pt]
\Lsq_{01}+\Lsq_{02}-\Lsq_{12} &= 2\guv y^u_1 y^v_2
\end{align*}
The solution is readily found to be $A=L_{01}$, $B=(\Lsq_{01}+\Lsq_{02}-\Lsq_{12})/(2L_{01})$
and $C=(\Lsq_{02}-B^2)^{1/2}$.

% --------------------------------------------------------------------------------------------
\subsection{The tetrad coordinates}
\label{sec:TetradCoords}

The transformation from the lattice to tetrad coordinates is quite simple. Let $e_a$ be the
basis for the tetrad frame and let $\partial_\mu$ be the corresponding basis for the lattice
frame. Recall that we have previously chosen the frames so that both $e_1$ and $\partial_t$
are aligned with the normal to the Cauchy surface. Now consider a typical vector $v_a$ that
joins $(0)$ to $(a)$. In the lattice frame this vector has components $y^\mu_a$ while in the
tetrad frame, with basis $e_b$, its components are just $v_a{}^b$. That is we have, for
$a=1,2,3$
\begin{gather}
n = \partial_t = e_1\\[3pt]
y^t_a = v_a{}^t\\[3pt]
v_a = y^\mu_a\partial_\mu = v_a{}^b e_b
\end{gather}
In the last equation both the $y^\mu_a$ and $v_a{}^b$ are known. Thus we have sufficient
information to compute $\partial_\mu$ in terms of $e_a$ and vice versa. Note that
the tetrad coordinates $x^\mu_a$ are given by
\begin{equation}
x^\mu_a = v_a{}^b \emb\>\quad {\rm with}\quad \emb=\delta^\mu{}_b
\end{equation}

Finally, using equation (\ref{eqn:TimeCoord}), we can compute the time coordinate for every
vertex, not just the three vertices associated with $v_a{}^b$, $a=1,2,3$
\begin{equation}
y^t_a = x^t_a = -\frac{1}{2}\Kmn x^\mu_a x^\nu_a\>,\quad a=1,2,3,\cdots
\end{equation}

% ============================================================================================
\section{Source terms}
\label{sec:SourceTerms}

We have previously mentioned, without giving details, that source terms such as $\dRxyxyz$ can
be computed by applying a finite difference approximation to data imported from neighbouring
cells. Here we will outline how such a procedure can be applied (the exact details will of
course depend on the structure of the lattice). The same procedure can also be used to
estimate the spatial derivatives of the $\ema$.

\begin{figure}[t]
\centerline{\includegraphics[width=\textwidth]{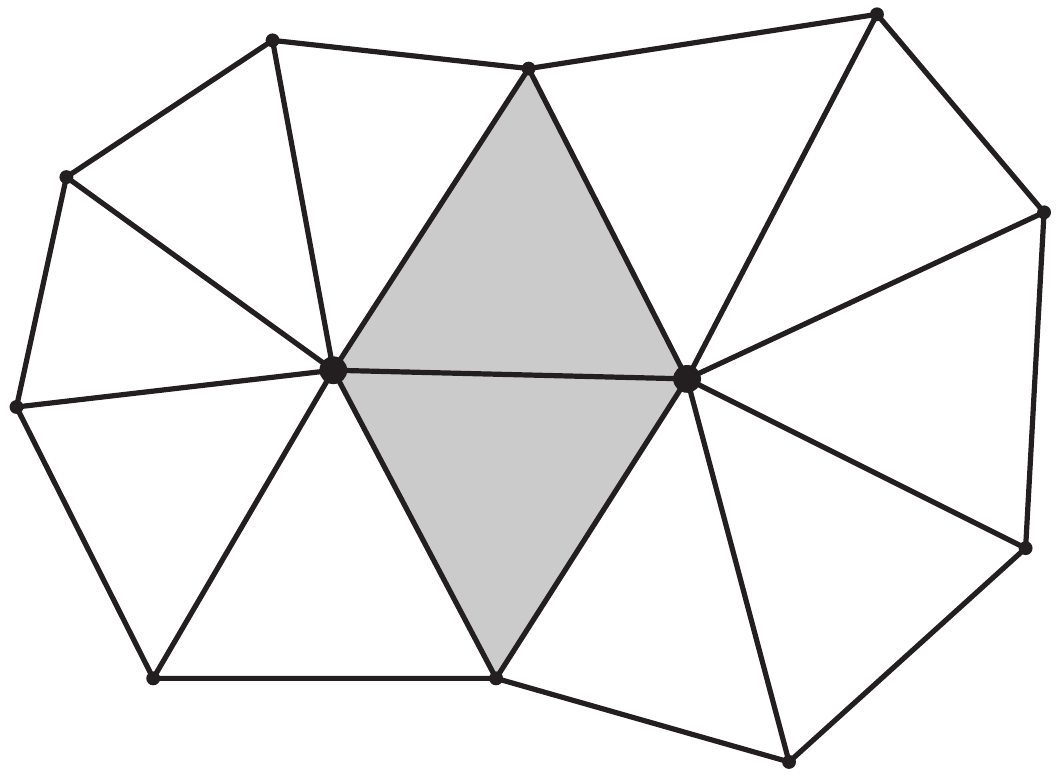}}
\vskip0.5cm
\caption{\normalfont An example of the overlap, the shaded region, between a pair of
computational cells. The central vertex of each computational cell is denoted by the large
dots whereas the smaller dotes denote the vertices that define the boundary of the computation
cells. These vertices are themselves the central vertices of other computational cells. In
this 2-dimensional example the overlap consists of just the pair of triangles. In 3 dimensions
the over lap would consist of a closed loop of tetrahedra. In each case there is ample
information available to obtain a coordinate transformation between the pair of local Riemann
normal frames.}
\label{img:TwoCells}
\end{figure}

Suppose we have two neighbouring computational cells that have a non-trivial overlap (as
indicated in Figure (\ref{img:TwoCells})). Each cell will carry values for $\Rxyxy$ in their
own local RNC frames. Our first task would be to import the values form the one cell to the
other. This will entail a coordinate transformation, composed of a boost (to account for the
change in the unit normal between the two cells) and a spatial rotation (to account for the
different orientations of the legs of the cells).

Let $x^\mu$ be the (tetrad) coordinates in one cell and let $x'^\mu$ be coordinates in the
other cell. Our plan is to import data form the $x'^\mu$ frame to the $x^\mu$ frame. We will
demand that the overlap region be such that it contains at least one set of three linearly
independent vectors (\ie legs), at $O'$, which we will denote by $w_i$, $i=1,2,3$. Since we
know the coordinates of each vertex in each cell we can easily compute the components of $w_i$,
$i=1,2,3$ in each frame. The normal vector $n_{o'}$ at $O'$ will have components $n'^\mu_{o'}
= (1,0,0,0)^\mu$ in the $x'^\mu$ frame. But in the $x^\mu$ frame we expect $n^\mu_{o'} =
n^\mu_o - K^\mu{}_\nu x^\nu_{o'}$. Thus we have 4 linearly independent vectors at $O'$,
expressed in two different frames, and so there must exist a mapping from the components in
one frame to those in the other. That is there exists a $U^\mu{}_\nu$ such that
\begin{align}
n^\mu_{o'} &= U^\mu{}_\nu n'^\nu_{o'}\\[3pt]
w^\mu_i &= U^\mu{}_\nu w'^\nu_i\>,\quad i=1,2,3
\end{align}
Since we have values for the components of $n_{o'}$ and $w_i$, $i=1,2,3$ in both frames we can
treat this as a system of equations for the $U^\mu{}_\nu$.

With the $U^\mu{}_\nu$ in hand, we can compute the values of $\Rmanb$ at $O'$ in the $x^\mu$
frame of $O$ by way of
\begin{equation}
\left(\Rmanb\right)_{o'} = 
     U_\mu{}^\theta U_\nu{}^\phi U_\alpha{}^\rho U_\beta{}^\tau 
     \left(R'_{\theta\phi\rho\tau}\right)_{o'}
\end{equation}
with $U_\mu{}^\nu = g_{\mu\alpha} g^{\nu\beta} U^\alpha{}_\beta$ and $g_{\mu\nu} =
\diag(-1,1,1,1)$. This can be repeated for all of the vertices that surround $O$. The result
is a set of point estimates for $\Rmanb$ in the neighbourhood of $O$ which in turn can be used
to estimate the derivatives of $\Rmanb$ at $O$. This part of the process is similar to that
required when computing the Hessian (see below) and presumably similar methods could be
applied.

Note that for a sufficiently refined lattice, the $U^\mu{}_\nu$ should be close to the
identity map, that is $U^\mu{}_\nu = \delta^\mu{}_\nu + V^\mu{}_\nu\BigO{L}$ where the
$V^\mu{}_\nu$ are each of order $\BigO{1}$. This can be used to simplify some of the above
computations.

See \cite{brewin:2010-02} for a complete example in the context of the Schwarzschild
spacetime.

In section (\ref{sec:NewRiemEvolve}) we noted that substitutions such as
$\Rtyxy=\rabcd\Eta\Eyb\Exc\Eyd$ could be introduced into the curvature evolution equation
(\ref{eqn:NewEvolve}). At that time we argued that that was not necessary for the coordinate
data, in this instance $\Rtyxy$, could easily be recovered when needed by using
$\Rtyxy=\rabcd\Eta\Eyb\Exc\Eyd$. Then the scheme described above could be used to compute
$\dRtyxyx$. However there may be numerical advantages in making a formal substitution before
estimating any of the partial derivatives. For $\dRtyxyx$ this would lead to the following
\begin{align*}
\dRtyxyx &= \left(\rabcd\Eta\Eyb\Exc\Eyd\right)_{,x}\\[3pt]
         &= \drabcdx\Eta\Eyb\Exc\Eyd + \rabcd\left(\Eta\Eyb\Exc\Eyd\right)_{,x}
\end{align*}
Since the $\rabcd$ are scalars, their partial derivatives can be estimated without requiring
any of the frame transformations described above (importing such data from neighbouring cells
is trivial). This leaves us with the derivatives of the form $(e_\mu{}^a)_{,x}$. Since $n_\mu
= -e_\mu{}^1$ we can use (\ref{eqn:DerivNormal}) to eliminate any of the spatial derivatives
of $e_\mu{}^1$, in this case $(e_\mu{}^1)_{,x}$. This would introduce the extrinsic curvatures
into the evolution equations. However the remaining partial derivatives, $(e_\mu{}^i)_{,x}$,
$i=2,3,4$, would have to be estimated using the methods described above (by importing data
from neighbouring cells etc.). This approach does incur a small computational overhead which
may be justified if it brings some improvement to the quality of the numerical data (\eg
better accuracy and or stability). Judging the merits of this variation against the simple
method given in section (\ref{sec:NewRiemEvolve}) might best be decided by direct numerical
experimentation.

% --------------------------------------------------------------------------------------------
\subsection{Extrinsic curvatures}
\label{sec:Extrinsic}

A cursory glance at equation (\ref{eqn:ADMDLija}) might give the impression that it
constitutes a simple linear system for the $\Kuv$. But things are never as simple as they
seem. The problem, as already noted, is that there are far too many equations for the six
$\Kuv$. If we make the reasonable assumption that the lattice data is a good approximation to
the (unknown) continuum spacetime then we can expect considerable redundancy in this
overdetermined system. How then do we pull out just six equations for the six $\Kuv$? One
option is to reject all but six of the equations and hope that this yields an invertible
system for the $\Kuv$. A better, and more flexible approach, is to take a weighted sum of the
equations, that is we create a new set of equations of the form
\begin{equation}
0 = \sum_{ab} W^n_{ab}\left(P_{ab} - \Kuv \Delta x^u_{ab} \Delta x^v_{ab}\right)
\end{equation}
where $W^n_{ab}$ are a set of weights of our own choosing (typical values being 0 and $\pm1$).
With $n=1,2,3\dots6$ we have six equations for the six unknowns. This idea has been used
previously \cite{brewin:1998-02} and worked very well. There are certainly other options that
could be explored (\eg different choices of weights, least squares) but we have tested none
simply because the above scheme seems to work well.

% --------------------------------------------------------------------------------------------
\subsection{The Hessian}
\label{sec:Hessian}

At some point we will need to estimate the $N_{\vert uv}$ at a central vertex. Since $N$ is a
scalar function and since we are using Riemann normal coordinates this computation is
essentially that of computing all of the second partial derivatives on an unstructured grid.
There is an extensive literature on this point in the context of finite element schemes. We
mention here one approach which we discussed in one of our earlier papers
\cite{brewin:1998-02} (but which we have yet to test).

Consider a typical leg $(ij)$ in some computational cell. We can estimate $N_{\vert u}$ at the
centre of the leg by the centred finite difference approximation
\begin{equation}
\left(N_{\vert u}\right)_{ij} = \frac{N_j-N_i}{\Lij} \left(m_u\right)_{ij}
\label{eqn:GradLapse}
\end{equation}
in which $\left(m_u\right)_{ij}$ is the unit vector tangent to the geodesic and oriented so
that it points from $(i)$ to $(j)$. We can repeat this computation for each leg in the
computational cell and then estimate $N_{\vert uv}$ by a least squares fit of the function
\begin{equation}
{\tilde N}_{\vert u}(x) = {\tilde N}_{\vert u} + {\tilde N}_{\vert uv} x^v
\end{equation}
to the data generated above by equation (\ref{eqn:GradLapse}). A suitable least squares sum
would be
\begin{equation}
S({\tilde N}_{\vert u},{\tilde N}_{\vert uv})
   = \sum_u \sum_{ij} 
     \left( \left(N_{\vert u}\right)_{ij}
            - {\tilde N}_{\vert u} 
            - {\tilde N}_{\vert uv}\> {\bar x}^v_{ij} \right)^2
\end{equation}
where ${\bar x}^v_{ij}$ is the centre of the leg $(ij)$. Note that this least squares fit must
be made subject to the constraint $N_{\vert uv}=N_{\vert vu}$. The coefficients ${\tilde
N}_{\vert u}$ and ${\tilde N}_{\vert uv}$ would then be taken as our estimates for the
corresponding quantities at the central vertex.

% ============================================================================================
\section{Discussion}
\label{sec:Discuss}

There are a number of aspects of this paper that could easily be debated. For example, should
we proceed with the substitutions such as $\Rtyxy=\rabcd\Eta\Eyb\Exc\Eyd$ in equation
(\ref{eqn:NewEvolve})? As already noted in section (\ref{sec:SourceTerms}) this would
introduce a raft of new terms including the extrinsic curvatures. We chose not to use the
substitution solely for reasons of simplicity. There is also a question over our choice of
tetrad. Do we really need to demand that the tetrad be orthonormal? Not at all. We could
choose to tie the tetrad to the legs of the lattice (and then the tetrad would no longer be
needed) but that would produce a coupling amongst all of the evolution equations (\eg the
evolution equation for $\rxyxy$ would be a linear combination of all of the evolution
equations for $\Rmanb$). The resulting equations would not be anywhere near as simple as those
listed in the Appendix. Then we have the issue of estimating partial derivatives on an
irregular lattice (for the Hessian and the source terms in the curvature evolution equations).
This is non-trivial but at least there is an extensive literature on the subject and so a
workable solution should not be too hard to find (which may be the least squares method
suggested in section (\ref{sec:Hessian})). All of these issues (and most likely others) can be
explored by direct numerical exploration on a non-trivial $3+1$ spacetime. We plan to report
on such investigations soon. For a simple application to the $1+1$ Schwarzschild spacetime
see \cite{brewin:2010-02}.

% ============================================================================================
\appendix

\section{The curvature evolution equations}

Here we list all 14 curvature evolution equations (this follows on from section
(\ref{sec:NewRiemEvolve}) where we provided details of the derivation for the first equation
below).
\bgroup
\def\m{\phantom{-}}
\def\v{\noalign{\vskip3pt\vskip 0pt plus 10pt\penalty-250\vskip 0pt plus-10pt}}
\begin{align}
\DrxyxyDt &= \dRtyxyx - \dRtxxyy\notag\\
          &\quad + \rtyxy\dNx - \rtxxy\dNy + \rtyxy\dNx - \rtxxy\dNy\label{eqn:riemEvolA}\\\v
\DrxyxzDt &= \m  \dRtzxyx - \dRtxxyz\notag\\
          &\quad + \rtyxz\dNx - \rtxxz\dNy + \rtzxy\dNx - \rtxxy\dNz\label{eqn:riemEvolB}\\\v
\DrxyyzDt &= \m  \dRtzxyy - \dRtyxyz\notag\\
          &\quad + \rtyyz\dNx - \rtxyz\dNy + \rtzxy\dNy - \rtyxy\dNz\label{eqn:riemEvolC}\\\v
\DrxzxzDt &= \m  \dRtzxzx - \dRtxxzz\notag\\
          &\quad + \rtzxz\dNx - \rtxxz\dNz + \rtzxz\dNx - \rtxxz\dNz\label{eqn:riemEvolD}\\\v
\DrxzyzDt &= \m  \dRtzxzy - \dRtyxzz\notag\\
          &\quad + \rtzyz\dNx - \rtxyz\dNz + \rtzxz\dNy - \rtyxz\dNz\label{eqn:riemEvolE}\\\v
\DryzyzDt &= \m  \dRtzyzy - \dRtyyzz\notag\\
          &\quad + \rtzyz\dNy - \rtyyz\dNz + \rtzyz\dNy - \rtyyz\dNz\label{eqn:riemEvolF}\\\v
\DrtyxyDt &= \m  \dRxyxyx - \dRxyyzz\notag\\
          &\quad + \riyxy\dNi + \rtyty\dNx - \rtxty\dNy\label{eqn:riemEvolG}\\\v
\DrtxxyDt &= -   \dRxyxyy - \dRxyxzz\notag\\
          &\quad + \rixxy\dNi + \rtxty\dNx - \rtxtx\dNy\label{eqn:riemEvolH}\\\v
\DrtzxyDt &= \m  \dRxyxzx + \dRxyyzy\notag\\
          &\quad + \rizxy\dNi + \rtytz\dNx - \rtxtz\dNy\label{eqn:riemEvolI}\\\v
\DrtzxzDt &= \m  \dRxzxzx + \dRxzyzy\notag\\
          &\quad + \rizxz\dNi + \rtztz\dNx - \rtxtz\dNz\label{eqn:riemEvolJ}\\\v
\DrtxxzDt &= -   \dRxyxzy - \dRxzxzz\notag\\
          &\quad + \rixxz\dNi + \rtxtz\dNx - \rtxtx\dNz\label{eqn:riemEvolK}\\\v
\DrtyxzDt &= \m  \dRxzxzx - \dRxzyzz\notag\\
          &\quad + \riyxz\dNi + \rtytz\dNx - \rtxty\dNz\label{eqn:riemEvolL}\\\v
\DrtzyzDt &= \m  \dRxzyzx + \dRyzyzy\notag\\
          &\quad + \rizyz\dNi + \rtztz\dNy - \rtytz\dNz\label{eqn:riemEvolM}\\\v
\DrtyyzDt &= \m  \dRxyyzx - \dRyzyzz\notag\\
          &\quad + \riyyz\dNi + \rtytz\dNy - \rtyty\dNz\label{eqn:riemEvolN}
\end{align}
\egroup

Note that in the above there are two instances of $\rtxyz$, in (\ref{eqn:riemEvolC}) and
(\ref{eqn:riemEvolE}), and these should be replaced with $\rtyxz-\rtzxy$.

% ============================================================================================
\section{Riemann normal coordinates}

We recall here a few basic properties of Riemann normal coordinates. A set of coordinates
$x^\mu$ are said to be in Riemann normal form if every geodesic passing through a given point
$O$ (the origin) is described by $x^\mu(s) = sv^\mu$ where $s$ is an affine parameter and
$v^\mu$ is constant along the geodesic. It follows from the geodesic equation and its
successive derivatives, that the connection and its higher symmetric derivatives\footnote{Here
we take a small liberty with notation, the upper index on the Christoffel symbol should be
ignored when computing covariant derivatives.} all vanish at the chosen point, that is at $O$
\begin{align}
0 &= \Gamma^\mu_{\alpha_1\alpha_2}\\
0 &= \Gamma^\mu_{(\alpha_1\alpha_2;\alpha_3\cdots\alpha_n)}\quad\quad n=3,4,5,\cdots
\end{align}
These conditions do not uniquely determine the coordinates for we are free to apply a
transformation of the form $x^\mu\mapsto\Lambda^\mu{}_\nu x^\nu$ which clearly preserves the
property that the geodesics through $O$ are of the form $x^\mu(s) = sv^\mu$. This freedom can
be used to ensure that the metric at $O$ is simply $g_{\mu\nu}=\diag(-1,1,1,1)$.

Choosing the coordinates so that the connection vanishes at the origin does introduce some
nice properties, in particular covariant differentiation reduces, at the origin, to simple
partial differentiation. This fact was essential to the analysis given in sections
(\ref{sec:EvolvRiemPt1}).

There are two main impediments to the existence of Riemann normal coordinates. The metric
must be smooth throughout the neighbourhood (\ie away from curvature singularities) and each
point in the neighbourhood should be connected to the origin by exactly one geodesic (\ie no
pair of geodesics through $O$ should cross, except at $O$). These conditions are easily
satisfied by simply choosing the neighbourhood around $O$ to be sufficiently small (but not
vanishingly small).

In these coordinates the metric and connection can be expanded as a Taylor series around $O$
leading to
\begin{align}
g_{\mu\nu}(x) &= g_{\mu\nu} 
               - \frac{1}{3} \Rmanb x^\alpha x^\beta 
               - \frac{1}{6} \dRmanbg x^\alpha x^\beta x^\gamma
               + \BigO{L^4}\\[5pt]
g^{\mu\nu}(x) &= g^{\mu\nu} 
               + \frac{1}{3} \upRmanb x^\alpha x^\beta 
               + \frac{1}{6} \updRmanbg x^\alpha x^\beta x^\gamma
               + \BigO{L^4}\\[5pt]
\Gamma^\mu_{\alpha\beta}(x) &= 
     \frac{1}{3} R^\mu{}_{\alpha\gamma\beta}x^\gamma
   + \frac{1}{24}\left(  2R^\mu{}_{\gamma\delta\beta,\alpha}
                       + 4R^\mu{}_{\alpha\delta\beta,\gamma}
                       +  R_{\gamma\alpha\delta\beta}{}^{,\mu}\right)x^\gamma x^\delta\notag\\
   &\quad + (\alpha \leftrightarrow \beta) 
   + \BigO{L^3}\label{eqn:RNCGamma}
\end{align}
If we know the Riemann normal coordinates, $x^\mu_i$ and $x^\mu_j$, for a pair of points, $i$
and $j$, then we can compute the length of the geodesic segment that joins the points by
\begin{equation}
\Lsqij = \left( g_{\mu\nu} 
            - \frac{1}{3} \Rmanb \Bar{x}^\alpha_{ij} \Bar{x}^\beta _{ij}
            - \frac{1}{6} \dRmanbg \Bar{x}^\alpha_{ij} \Bar{x}^\beta_{ij} \Bar{x}^\gamma_{ij} 
         \right) \Dxij^\mu \Dxij^\nu
         + \BigO{L^6}
\end{equation}
where $\Dxij^\mu:=x^\mu_j-x^\mu_i$ and $\Bar{x}^\mu_{ij}:=(x^\mu_j+x^\mu_i)/2$ is the
mid-point of the leg. The unit tangent vector $m^\mu_{ij}$ to the geodesic at $i$, is given by
\begin{align}
L_{ij} m^\mu_{ij}&=\Delta{x_{ij}^{\mu}} 
+\frac{1}{3} x^{\alpha}
            \Delta{x_{ij}^{\nu}}
            \Delta{x_{ij}^{\beta}} R^{\mu}{}_{\nu\alpha\beta} 
+\frac{1}{12} x^{\alpha} 
                x^{\nu}
                \Delta{x_{ij}^{\beta}}
                \Delta{x_{ij}^{\gamma}}
                {R^{\mu}{}_{\alpha\nu\gamma,\beta}}\notag\\
&\quad+\frac{1}{6} x^{\alpha} 
               x^{\nu}
               \Delta{x_{ij}^{\beta}}
               \Delta{x_{ij}^{\gamma}}
               {R^{\mu}{}_{\beta\nu\gamma,\alpha}} 
+\frac{1}{24} x^{\alpha} 
                x^{\nu}
                \Delta{x_{ij}^{\beta}}
                \Delta{x_{ij}^{\gamma}}
                {R_{\alpha\beta\nu\gamma}^{,\mu}}\\
&\quad+\frac{1}{12} x^{\alpha}
                \Delta{x_{ij}^{\nu}}
                \Delta{x_{ij}^{\beta}}
                \Delta{x_{ij}^{\gamma}}
                {R^{\mu}{}_{\beta\alpha\gamma,\nu}}\notag
\end{align}
Finally, if we have a geodesic triangle built on the three points $i$, $j$, $k$ then the
generalised cosine law takes the form
\begin{equation}
2\Lik\Ljk \cos\theta_{ij} = \Lsqik + \Lsqjk - \Lsqij
   - \frac{1}{3} \Rmanb\>
     \Delta x^\mu_{ik} \Delta x^\nu_{ik}
     \Delta x^\alpha_{jk} \Delta x^\beta_{jk} + \BigO{L^5}
\end{equation}
in which $\theta_{ij}$ is the angle subtended at vertex $k$ by the geodesic that connects $i$
to $j$. 

% ============================================================================================
\clearpage

\def\Figure#1#2{%
\centerline{%
\includegraphics[width=#1\textwidth]{#2}}
\vskip0.5cm}

\def\FigPair#1#2{%
\centerline{%
\includegraphics[width=0.6\textwidth]{#1}\hfill%
\includegraphics[width=0.6\textwidth]{#2}}
\vskip0.5cm}

\def\FigPairV#1#2{%
\centerline{%
\includegraphics[width=0.95\textwidth]{#1}}
\vskip0.25cm
\centerline{%
\includegraphics[width=0.95\textwidth]{#2}}
\vskip0.5cm}

\def\FigQuad#1#2#3#4{%
\centerline{%
\includegraphics[width=0.6\textwidth]{#1}\hfill%
\includegraphics[width=0.6\textwidth]{#2}}%
\centerline{%
\includegraphics[width=0.6\textwidth]{#3}\hfill%
\includegraphics[width=0.6\textwidth]{#4}}
\vskip0.5cm}

% === figures ================================================================================

% \begin{figure}[t]
% \centerline{\includegraphics[width=0.8\textwidth]{images/fig01}}
% \vskip0.5cm
% \caption{\normalfont This is a figure caption}
% \label{img:LamEta}
% \end{figure}
% 
% \begin{figure}[t]
% \centerline{\includegraphics[width=0.8\textwidth]{images/fig02}}
% \vskip0.5cm
% \caption{\normalfont And a second figure.}
% \label{img:Theta}
% \end{figure}
% 
% \begin{figure}[t]
% \FigPairV{plots/04}%
%          {plots/11}
% \caption{\normalfont%
% Some features of the Schwarzschild spacetime.}
% \label{fig:Schwarzschild}
% \end{figure}

\clearpage

% ============================================================================================
% \bibliographystyle{brewin}  % use this to create the bibliography
% \bibliography{brewin}

\providecommand{\href}[2]{#2}\begingroup\raggedright\endgroup
           % use this when submitting the paper to the journal

% If I use references like \PaperI (rather than a \cite{me}) then I need to manually edit
% the paper.bbl file (after running bibtex) as follows.

%% old item
%
% \bibitem{brewin:2002-01}
% L.~Brewin, {My important paper}, 
%            {\em CGG } {\bf 19} (2002) 1-25.
% 

%% new item
%
% \bibitem{brewin:2002-01}
% L.~Brewin, {\hypertarget{paper1}{{\bf (Paper 1)}} My important paper}, 
%            {\em CGG } {\bf 19} (2002) 1-25.
% 

\end{document}